\documentclass[%
 reprint,
 amsmath,amssymb,
 aps,
]{revtex4-1}

\usepackage{graphicx}% Include figure files
\usepackage{dcolumn}% Align table columns on decimal point
\usepackage{bm}% bold math
\def\be{\begin{equation}}
\def\ee{\end{equation}}
\def\ba{\begin{align}}
\def\ea{\end{align}}

\def\Pb{\mathcal{P}}

\def\intL{\int_0^L}
\def\ft{\tilde{f}}
\DeclareMathOperator{\erf}{erf}

\begin{document}

\title{Inequivalence of fixed-force and fixed-extension statistical ensembles for  a flexible polymer tethered to a planar substrate}

 \author{Sandipan Dutta} 
 \affiliation{Center for Soft and Living Matter, Institute for Basic Science, Ulsan 44919, Korea}
 \author{Panayotis Benetatos} 
\affiliation{Department of Physics, Kyungpook National University, 80 Daehakro, Bukgu, Daegu 41566, Korea.}
\email{ pben@knu.ac.kr }

\begin{abstract}
Recent advances in single macromolecule experiments have sparked interest in the ensemble dependence of force-extension relations. The thermodynamic limit may not be attainable for such systems, that leads to inequivalence of the fixed-force and the fixed-extension ensemble. We consider an ideal Gaussian chain described by the Edwards Hamiltonian with one end tethered to a rigid planar substrate. We analytically calculate the force-extension relation in the two ensembles and we show their inequivalence which is caused by the confinement of the polymer to half space. The inequivalence is quite remarkable for strong compressional forces. We also perform Monte-Carlo simulations of a tethered wormlike chain with contour length 20 times its persistence length which corresponds to experiments measuring the conformations of DNA tethered to a wall.  The simulations confirm the ensemble inequivalence and qualitatively agree with the analytical predictions of the Gaussian model. Our analysis shows that confinement due to tethering causes ensemble inequivalence, irrespective of the polymer model.  \\%The abstrast goes here instead of the text "The abstract should be..."
\end{abstract}

\maketitle

%Please use \dag to cite the ESI in the main text of the article.
%If you article does not have ESI please remove the the \dag symbol from the title and the footnotetext below.
%\footnotetext{\dag~Electronic Supplementary Information (ESI) available: [details of any supplementary information available should be included here]. See DOI: 10.1039/cXsm00000x/}
%additional addresses can be cited as above using the lower-case letters, c, d, e... If all authors are from the same address, no letter is required

%%%END OF FOOTNOTES%%%

%%%MAIN TEXT%%%%

\section{\label{sec:Introduction}Introduction}

 In order to measure the elasticity of polymers, which in turn provides information about their conformational properties, we need to perturb them with an external force or confine them with steric constraints. This is usually done on macroscopic systems consisting of many macromolecules, and then  information relevant to a single molecule is extracted indirectly. In recent years, however, it has become possible to perform measurements on a single molecule \cite{neuman2008single,strick2002stretching} using optical tweezers \cite{bustamanteoptical2008}, magnetic tweezers \cite{bustamantemagnetic1992,salehmagnetic2009}, atomic force microscopy (AFM) \cite{GaubAFMScience1997}, or other micro-mechanical methods (e.g., flow stretching \cite{ChuScience}). The interpretation of these experiments relies on the methods and concepts of thermodynamics and statistical mechanics. It was Paul Flory who was the first to point out the analogy between single-polymer force-extension relations and the pressure-volume equation of state of a gas, and their interpretation in the context of statistical ensembles (Gibbs and Helmholtz) \cite{Florybookchains}.  As opposed to the conventional macroscopic systems, however, the thermodynamics of single polymer systems can be very different. These peculiarities can be traced to the lack of a well-defined thermodynamic limit for many single-polymer systems. Despite having an enormous number of monomers, their strong interactions and low dimensionality, which imply strong fluctuations, can make their thermodynamic behavior sensitive to boundary conditions \cite{Skvortsovreview2012}. Apart from the force-extension relations, this sensitivity extends to phase transitions (e.g., escape transition  \cite{SkvortsovtranslocationJCP2007}  or desorption \cite{SkvortsovBinderPRE2012,JuttaBinderJCP2014}) which also depend on the statistical ensemble.

\begin{figure}
 \centering
 \includegraphics[width=0.5\textwidth]{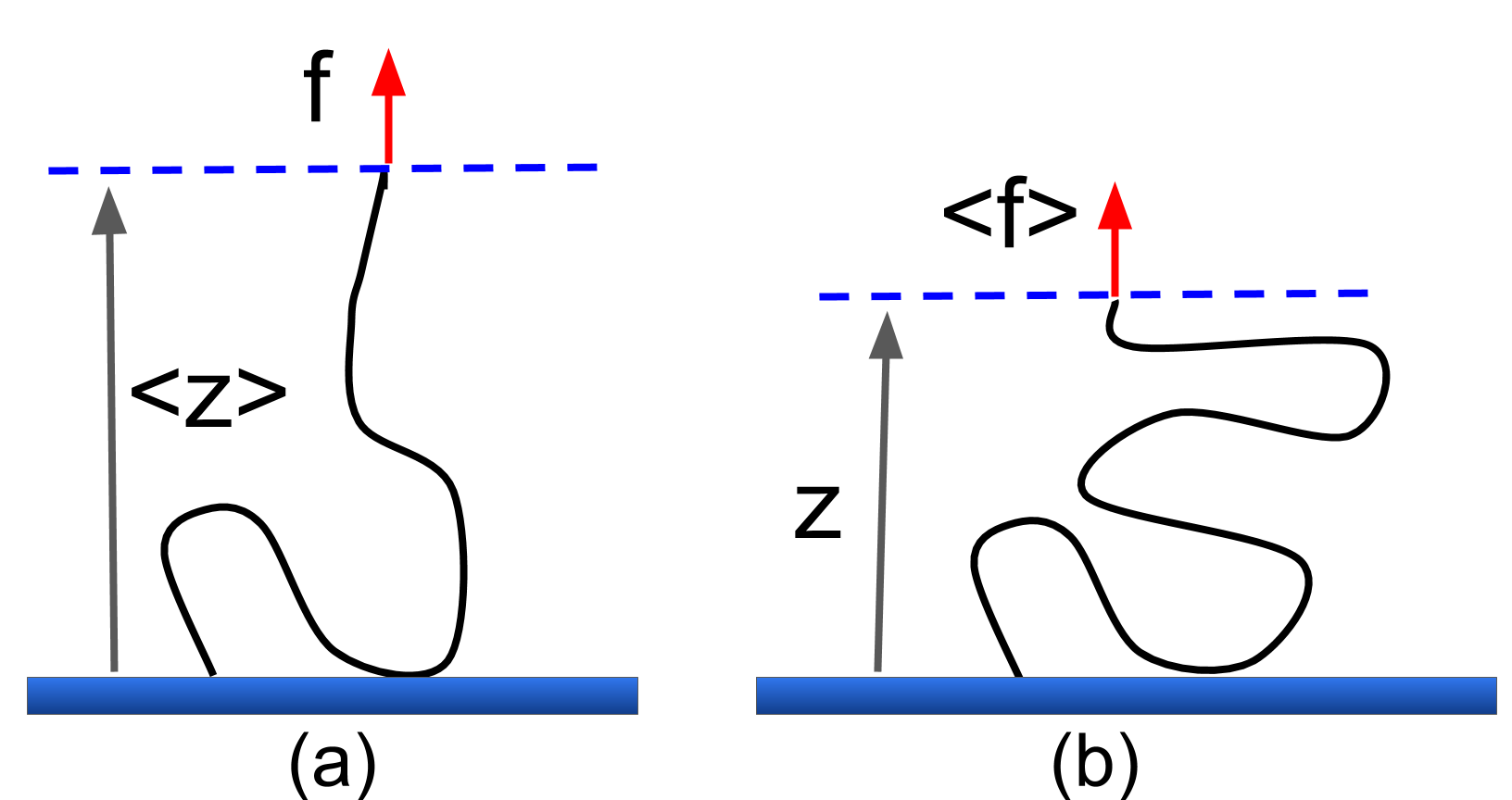}
 \caption{A schematic diagram of a polymer tethered to an impenetrable wall, (a) in the fixed-force and (b) in the  fixed-extension ensemble.}
 %A constant force $\mathbf{f}$ is applied to the free end of the polymer in the fixed-force ensemble also referred to as Gibbs ensemble and average extension $\langle z\rangle$ is measured. In fixed-extension ensemble also known as Helmholtz ensemble, the polymer has a constant extension while the average force is measured. 
 \label{fig:wall}
\end{figure}

The Gibbs ensemble refers to force-extension measurements of a polymer where a fixed force (tension)  is applied to its end which is free to fluctuate, while the Helmholtz ensemble is obtained when the extension of the polymer (given by its end-to-end distance) is fixed whereas the associated force (tension) fluctuates. In the former case, an average extension is extracted, whereas in the latter it is an average tension. These two ensembles are also referred to in the literature as isotensional and isometric or fixed-force and fixed-extension, respectively. It is known that in the thermodynamic limit, all ensembles yield the same equation of state \cite{Callen}. The thermodynamic limit of a physical system refers to the limit where its size is large enough so that boundary (finite-size) effects can be neglected and its bulk behavior is uniform. The uniform behavior becomes manifest when the equation of state is brought in a form that contains only intensive variables. In fluid systems, the number of particles and the volume are taken to infinity, keeping the density constant. Assuming short-range interactions between the particles, the thermodynamic limit exists. For example, both the ideal gas equation and the van der Waals equation can be written in a form that contains only the pressure, the temperature, and the density.  In continuous polymer systems, the size of the system as measured by the total contour length with respect to another characteristic length (such as the persistence length or the Kuhn length or, in the case of polymers under tension, $k_B T/f$ or $\sqrt{\kappa/f}$, where $\kappa$ is the bending siffness).  In the thermodynamic limit, that ratio is taken to infinity. At that large-size limit, we may get uniform behavior in the sense that linking sequentially together two polymers with the same intensive variables results in a polymer with exactly the same behavior.  For polymers under tension, the system is in the thermodynamic limit, if the resulting deformation is additive and proportional to the contour length. In polymers,  the (uniform) thermodynamic limit may not exist. This is clear in the case of semiflexible polymers. Monomers are strongly correlated to give rise to a finite bending rigidity and the related persistence length. The behavior of a semiflexible polymer is dominated by its persistence length which is of the order of the total contour length.  If, in order to reach a thermodynamic limit, we take the latter  to infinity keeping the former fixed, we get a flexible chain with manifestly different behavior. The non-existence of a thermodynamic limit for semiflexible polymers results in the inequivalence of the two ensembles which has been analyzed in \cite{SinhaSamuel,ChaudhuriPRE2007,winkler2003deformation,glatting1993partition,van2006one}.  Those studies have shown that, for persistence lengths of the order of the contour length, the free energies in the two ensembles are not related by a simple Legendre transformation and they yield qualitatively different force-extension curves.  More specifically, the fixed-extension ensemble exhibits non-monotonic force-extension curves which are not possible in the fixed-force ensemble. We should point out that semiflexible polymers are inherently   finite-size systems and different boundary conditions (e.g., free hinged-hinged, transverse-position-constrained hinged-hinged, clamped-clamped, etc.) yield different force-extension relations even within the Gibbs ensemble \cite{PBMotor,PBHinge}. For example, when we constrain the end-points of a wormlike chain with hinged-hinged boundary conditions to fluctuate on a straight line defined by the tension force, the average extension for a given force is greater than that in the case  of free hinged-hinged boundary conditions \cite{PBHinge}. The existence of a tension-dependent  correlation length (inversely proportional to the square root of the tension) \cite{PGDG_Ciferri,PBEMJ}  allows approach to a thermodynamic limit as it becomes much smaller than the total contour length at the strong-tension limit, and convergence of all force-extension relations to a single curve \cite{PBMotor,PBHinge}. 

As opposed to semiflexible polymers, the Gaussian chain modeled by the Edwards Hamiltonian \cite{DoiEdwards} is a fractal object and as such is expected to have a well-defined thermodynamic limit. Despite its simplicity (relative to models of semiflexible polymers) the existence of a thermodynamic limit and the equivalence of the two ensembles has been a matter of debate \cite{WinklerSoft,manca2014equivalence,SuzenPRE2009,NeumannBPJ2003,NeumannPRA1986,NeumannPRA1985}. The thermodynamic limit of a single polymer may become ill-defined because the polymer itself is dominated by its finite size, or because the polymer is somehow confined in space. The former case appears when the monomer interactions give rise to long-range correlations (e.g., orientational correlations in semiflexible polymers), and also in polymers with bistable elements (e.g., folded-unfolded) \cite{giordano2018helmholtz,manca2013two}.   In this work, we study a Gaussian chain with one end tethered to a rigid planar substrate (Gaussian mushroom) as shown in Fig. \ref{fig:wall}. The presence of the impenetrable substrate restricts the conformations of the chain and renders the thermodynamic limit problematic. Our goal is to calculate the force-extension relation in the two ensembles. The inequivalence of the ensembles for a tethered polymer on a hard surface with hard-core bead-bead interactions has been demonstrated numerically in \cite{JuttaBinderJCP2014}. The study applied Wang-Landau simulation techniques on the bond fluctuation (BF) model of a flexible chain. In our analysis of the Gaussian mushroom, we obtain analytic force-extension relations, in closed form,  for the two ensembles. Our calculation is based on the free end distribution of a Gaussian mushroom which is known to be a Rayleigh function \cite{Slutsky}.  Since the Gaussian  chain is believed not to exhibit ensemble inequivalence in free space, it is the ideal polymer model for the investigation of the effect of confinement due to tethering on the ensemble equivalence.

We should point out that, apart from the theoretical motivation described above, the tethered configuration of a biopolymer is very commonly used in single-molecule manipulation experiments \cite{tetheredDNA_AJP,SegallPNelsonPRL2006,MeinersPRE(R)2009}. It is also relevant to biological processes in living cells. Surface-tethered adhesion receptors play an important role in the control of the interaction between cells and substrates or bacteria and tissues \cite{BiomimeticDillowbook}. This geometry also affects biologically relevant ligand-receptor kinetics \cite{Terentjev-tethered}.

In addition to the analytic study of the Gaussian mushroom, we also perform Monte-Carlo simulations of the wormlike chain (WLC) model tethered to a wall with parameters appropriate for a short DNA chain. Our motivation is the recent experiment work by Lindner et al.  \cite{Lindner} which measured the three-dimensional distribution of the free end of a tethered DNA molecule. That experiment found an axial distribution which qualitatively is Rayleigh-like and agrees with simulations of the WLC model. The transverse (in-plane) distribution was found to be Gaussian. The deviation of the axial distribution from the Rayleigh function for relatively long chains was a significant result of that study and was explained by the increased number of collisions of the long chain with the substrate. Collisions would have  a different effect in the WLC model from that in the Gaussian model, because of the finite bending rigidity of the former. Our simulations aim at predicting the outcome of a modified version of the experiments of Lindner et al. in which the free end is pulled (or pushed) by a force. Our results confirm that confinement due to the tethering substrate causes ensemble inequivalence, irrespective of whether we use the Gaussian chain model, the WLC model, or the BF model of Ref. \cite{JuttaBinderJCP2014}.

Our article is organized as follows. In Sec. 2, we review the equivalence of the two ensembles for a Gaussian chain modeled by the Edwards Hamiltonian in free space, by calculating the corresponding force-extension relations. In Sec. 3, we analytically calculate the average-force versus extension relation in the isometric ensemble from the conformational probability distribution which has a Rayleigh form. We demonstrate the lack of a thermodynamic limit and its recovery at appropriate limits of the force or the number of monomers. In Sec. 4, we do the same for the average-extension versus force relation in the isotensional ensemble. We show where the two ensembles differ and the limits where they converge. In Sec. 5, we present the analytical calculation of the differential compliance in the two ensembles, which is an important measure of the tensile or compressional elasticity of our chain. The force-extension relations in the two ensembles are computed by Monte Carlo simulations for a WLC with parameters corresponding to the experiment by Lindner et al.\cite{Lindner} and the results are shown in Sec. 6. We summarize and conclude in Sec. 7. In the Appendix, we discuss some details of the simulation, and we show the corresponding conformational probability distribution.

\section{Equivalence of ensembles for a Gaussian chain in free space}

We model a flexible polymer as a Gaussian chain whose one end is fixed at the origin of the coordinate system. Its partition function is given by the functional integral:\\
\begin{equation}
{\cal Z}_0({\bf r})={\cal N}\int_{{\bf r}(0)=0}^{{\bf r}(L)={\bf r}}{\cal D}[\{{\bf r}(s)\}]\exp\big(-\beta {\cal H}_0({\bf r}(s))\big)\;,
\end{equation}
where the functional integration is over all random paths between $0$ and ${\bf r}$, and ${\cal H}_0$ is the Edwards Hamiltonian \cite{DoiEdwards},\\
\begin{equation}
{\cal H}_0=\frac{3}{4l_p\beta}\intL\left(\frac{\partial{\mathbf{r}(s)}}{\partial s}\right)^2ds\;,
\end{equation}
${\cal N}$ is a normalization prefactor (which can be incorporated in the measure \cite{Kleinert}), $\beta=1/k_B T$,  $l_p$ is the persistence length (equal to half the Kuhn length), and $L$ is the contour length (related to the degree of polymerization, $N$, by $L=2l_pN$). Fixing one end excludes translation of the entire chain.

The end-to-end distance follows a Gaussian distribution \cite{DoiEdwards},\\
\begin{equation}
P_0({\bf r})=\Big(\frac{3}{4\pi L l_p}\Big)^{3/2}\exp\Big(-\frac{3 \mathbf{r}^2}{4Ll_p}\Big)\;,
\end{equation}
and $P_0({\bf r})\sim \Omega({\bf r})$, where $\Omega({\bf r})$ is the number of conformations with end-to-end vector ${\bf r}$. The entropy is given by Boltzmann's formula, $S=k_B\ln \Omega$, and the Helmholtz free energy is $F=U-TS$. Since there are no interactions in the ideal Gaussian chain, the only ${\bf r}$-dependence of the free energy comes from the entropy. The associated (conjugate) 
entropic force is\\
\begin{equation}
\langle {\bf f} \rangle =\nabla_{\bf r}F=\frac{3k_B T}{2 L l_p}{\bf r}\;.
\label{fext_free_isometric}
\end{equation}
This is the well known result of the ideal chain as entropic spring \cite{deGennesScaling}.

In the above-mentioned calculation, the extension is fixed and the average force is calculated. When we apply a finite tension ${\bf f}$ to the chain, the Hamiltonian acquires an interaction term:\\
\begin{equation}
{\cal H}={\cal H}_0-{\bf f}\cdot {\bf r}={\cal H}_0-\int_0^L {\bf f}\cdot \Big(\frac{\partial {\mathbf{r}(s)}}{\partial s}\Big)\; ds\;.
\end{equation}
The average extension under the tension ${\bf f}$ is
\begin{align}
\langle {\bf r}\rangle &=
\frac{\int_{{\bf r}(0)=0} {\cal D}[\{{\bf r}(s)\}] {\bf r}\exp\Big(-\frac{3}{4l_p}\intL\left(\frac{\partial{\bf r}}{\partial s}\right)^2ds+\beta \int_0^L{\bf f}\cdot \Big(\frac{\partial {\bf r}}{\partial s}\Big)ds\Big)}{\int_{{\bf r}(0)=0} {\cal D}[\{{\bf r}(s)\}] \exp\Big(-\frac{3}{4l_p}\intL\left(\frac{\partial{\bf r}}{\partial s}\right)^2ds+\beta \int_0^L{\bf f}\cdot \Big(\frac{\partial {\bf r}}{\partial s}\Big)ds\Big)} \nonumber\\
& =\frac{\int_{{\bf r}(0)=0} {\cal D}[\{{\bf r}(s)\}] {\bf r}\exp\Big(-\frac{3}{4l_p}\intL\left(\frac{\partial{\bf r}}{\partial s}-\frac{2l_p \beta}{3}{\bf f}\right)^2ds \Big)}{\int_{{\bf r}(0)=0} {\cal D}[\{{\bf r}(s)\}] \exp\Big(-\frac{3}{4l_p}\intL\left(\frac{\partial{\bf r}}{\partial s}-\frac{2l_p \beta}{3}{\bf f}\right)^2ds\Big)}\nonumber\\
&=\frac{\int_{{\bf R}(0)=0} {\cal D}[\{{\bf R}(s)\}]\big ({\bf R}+\frac{2l_p\beta}{3}{\bf f}L\big)\exp\Big(-\frac{3}{4l_p}\intL\left(\frac{\partial{\bf R}}{\partial s}\right)^2ds \Big)}{\int_{{\bf R}(0)=0} {\cal D}[\{{\bf R}(s)\}] \exp\Big(-\frac{3}{4l_p}\intL\left(\frac{\partial{\bf R}}{\partial s}\right)^2ds\Big)} \nonumber \\
& =\frac{2Ll_p}{3k_B T}{\bf f}\;.
\label{fext_free_isotensional}
\end{align}

The identity of the two force-extension relations, Eq. (\ref{fext_free_isometric}) and Eq. (\ref{fext_free_isotensional}),  establishes the equivalence of  the isometric (fixed extension, Helmholtz) and isotensional (fixed force, Gibbs) ensembles for the free ideal chain. 
%We point out that, in the above-mentioned calculations on the continuous Gaussian chain, the thermodynamic limit ($N\rightarrow\infty$) is related to the invariance of all observables to the rescaling $N\rightarrow \lambda N$ and $l_p\rightarrow \frac{1}{\sqrt{\lambda}}l_p$.

The equivalence or inequivalence of ensembles (in the thermodynamic limit) for several models of flexible polymers in unbounded space has been the subject of debate in the literature \cite{WinklerSoft}.  Winkler \cite{WinklerSoft} and also  Manca et al. \cite{manca2014equivalence} have argued that the ensemble inequivalence for flexible chains claimed by previous authors can be traced to the  inappropriate choice of pairs of conjugate variables (the average of the modulus of a fluctuating vector, in general, differs from the modulus of the average of the vector). That debate is beyond the scope of the present paper. Here, we focus on the effect of confinement due to tethering an ideal chain to a planar substrate and use the continuous (Edwards) model of the ideal Gaussian chain.

\section{Isometric ensemble}

In the rest of the paper, we consider a Gaussian chain with one end tethered to a planar substrate which acts as a hard wall. We choose a coordinate system such that the origin is at the tethering point and the chain is confined in the $z>0$-plane, as shown in Fig. \ref{fig:wall}. From the Edwards Hamiltonian, it is clear that the three dimensions decouple. The substrate leaves the $xy$-projection of the chain unaffected and it only affects the $z$-direction. 

It is known that the the probability density of the height of the free end of a tethered 
Gaussian polymer is given by the Rayleigh distribution \cite{Slutsky,Lindner}:\\
\be
\label{Rayleigh}
\Pb(z) = \frac{3 z}{2L l_p}\exp(-3z^2/4L l_p),\;\;\;z>0.
\ee
There are several ways one can obtain this Rayleigh distribution. One way is the method of images. The probability distribution of a Gaussian chain satisfies a diffusion equation with the arc-length parameter playing the role of time. The solution in the presence of the substrate (which introduces an absorbing boundary condition requiring the probability to vanish at $z=0$) involves a source and an image sink whose relative distance is taken to zero. The unnormalized solution of the diffusion equation for a chain originating a small distance $a$ above the substrate is the superposition of a Gaussian centered at $z=a$ and a "mirror image" Gaussian with a negative sign centered at $z=-a$
\begin{widetext}
\begin{eqnarray}
\label{images}
&\Pb_a(z) \propto \Big(\frac{3}{4\pi L l_p}\Big)^{1/2}\Big(\exp\Big(-\frac{3 (z-a)^2}{4Ll_p}\Big)-\exp\Big(-\frac{3 (z+a)^2}{4Ll_p}\Big)\Big)&\nonumber\\
&\propto\Big(\frac{3}{4\pi L l_p}\Big)^{1/2}\exp\Big(-\frac{3 (a^2+z^2)}{4Ll_p}\Big)\Big(\exp\Big(\frac{3 za}{2Ll_p}\Big)-\exp\Big(-\frac{3 za}{Ll_p}\Big)\Big)&\nonumber\\
&\propto\Big(\frac{3}{4\pi L l_p}\Big)^{1/2}\exp\Big(-\frac{3 (a^2+z^2)}{4Ll_p}\Big)\frac{3za}{Ll_p}\;,
\end{eqnarray}
\end{widetext}
where in the last line we keep the leading order contribution from the two exponentials for small $a$. If we normalize the distribution before we take the limit $a\rightarrow 0$, we obtain Eq. (\ref{Rayleigh}). Another way was suggested by Chandrasekhar \cite{chandrasekhar1943stochastic}, who counted the number of random walks reflecting on the substrate using a reflection principle. Slutsky \cite{Slutsky},  using functional integration, explicitly calculated the Fourier transform of the partition function of a Gaussian chain tethered to a delta-function planar potential. He extracted the Rayleigh distribution by taking the amplitude of the potential to infinity. 

The substrate acts as a geometric constraint, and all the conformations of the tethered chain with $z>0$ have the same energy. Therefore, the Helmholtz free energy, $F(z)$, depends on $z$ only through the entropy:
\begin{align}
 \beta F(z) = \gamma^2z^2-\ln(\gamma z) + \text{ const}\;,
\end{align}
where $\gamma = \sqrt{\frac{3}{4Ll_p}}$.

Thus the average entropic force is
\begin{equation}
\label{entropicforceGaussian}
\langle f \rangle = \frac{\partial F}{\partial z} =k_B T (2\gamma^2 z - \frac{1}{z}) \;,
\end{equation}
which can be written as
\be
\label{force-extension-Helmholtz}
\langle \tilde{f} \rangle = 2\tilde{z} - 1/\tilde{z}\;,
\ee
with $\tilde{f} = \beta f/\gamma$ and $\tilde{z}=\gamma z$. We notice that, for $\tilde{z}\gg1$, we recover the force-extension relation of the free  (untethered) Gaussian chain, 
\begin{align}
\label{force-extension-Helmholtz-tensile}
\langle f \rangle = \frac{3 k_B T}{2 l_p} \frac{z}{L}\;.
\end{align}

In the strongly compressional regime, the second term in the right hand side of Eq. (\ref{force-extension-Helmholtz}) dominates over the first, and we obtain a force-extension relation which does not depend on the size of the polymer  which is a feature that applies to the thermodynamic limit,
\begin{align}
\label{force-extension-Helmholtz-compress}
\langle f \rangle =-\frac{k_BT}{z}\;.
\end{align}
The divergence of the entropic force in the limit of vanishing axial extension is a result of the vanishing of the conformational probability to find the free end at $z=0$. Even though Eq. (\ref{force-extension-Helmholtz-compress}) does not contain the size of the system, it can be rewritten as a force-relative extension relationship which clearly depends on the system size,
\begin{align}
\label{force-strain-Helmholtz-compress}
 \frac{z}{L} =-\frac{k_BT}{L \langle f \rangle}\;.
\end{align}
We see that, in the compressive regime, it is impossible to express the force-extension relation in terms of purely intensive (relative extension, force) variables for $L\rightarrow \infty$. In that limit, the relative extension vanishes for any negative force.  Thus, the only bona fide thermodynamic limit of a tethered Gaussian chain occurs in the tensile ($f>0$) regime.

\section{Isotensional ensemble}

The partition function of a Gaussian chain tethered to a planar delta-function potential at one end ($s=0$) and subject to a force ${\bf f}$ exerted on the other end ($s=L$) reads (up to a normalization prefactor)
\begin{equation}
\label{isotensional_Z}
{\cal Z}_f = \int_{{\bf r}(0)=0}{\cal D}[\{{\bf r}(s)\}]\exp\big(-\beta \big({\cal H}_0+{\cal H}_f+{\cal H}_g\big)\big)\;,
\end{equation}
where ${\cal H}_0({\bf r}(s))$ is the Edwards Hamiltonian, 
\begin{equation}
{\cal H}_f({\bf r}(s))=-{\bf f}\cdot \big({\bf r}(L)-{\bf r}(0)\big)\;,
\end{equation}
and ${\cal H}_g({\bf r}(s))$ is the wall potential
\begin{equation}
{\cal H}_g({\bf r}(s))=g\int_0^L\delta[z(s)-z(0)]ds\;.
\end{equation}
At the limit $g\rightarrow \infty$, the $z=0$ plane becomes impenetrable and the partition function describes a Gaussian chain constrained to fluctuate in the $z>0$ or in the $z<0$ half-space, pulled away or pushed towards the $z=0$ plane by the force ${\bf f}=f\hat{\bf z}$, respectively (assuming $f>0$).
    
As in the previous sections, the transverse directions ($xy$) are unaffected by the force and the substrate, and can be ignored. The functional integral of Eq. (\ref{isotensional_Z}) can be calculated in two steps. At first, we integrate over all conformations from $z(0)=0$ to $z(L)=z$. Then, we integrate over $z$. The first integral has been calculated by Slutsky \cite{Slutsky} and, at the $g\rightarrow \infty$ limit, yields the Rayleigh distribution. Thus the problem reduces to a simple integration:
\begin{widetext}
\begin{align}
\label{isotensional_Z_1}
{\cal Z}_f &=\lim_{g \to \infty}\int_0^{\infty}dz \exp(\beta fz)\int_{z(0)=0}^{z(L)=z}{\cal D}[\{z(s)\}]\exp\big(-\beta \big({\cal H}_0+{\cal H}_g\big)\big)\nonumber\\
&= \int_0^{\infty}dz \exp(\beta fz)\frac{3 z}{2L l_p}\exp(-3z^2/4L l_p)\nonumber\\
&= \frac{1}{4}\left(2+\exp(\ft^2/4)\ft\sqrt{\pi}(1+\erf(\ft/2))\right),
\end{align}
\end{widetext}
where, in the last line, we use the rescaled and dimensionless force $\ft$.

From the partition function at constant tension $f$, we calculate the relevant free energy (Gibbs free energy), $G(\ft)=-k_B T\ln({\cal Z}_f )$.  The force extension relation is obtained by taking the derivative of the Gibbs free energy,
\begin{align}
\label{force-extension-Gibbs}
\langle {\tilde z} \rangle & = -\frac{\partial G(\ft)}{\partial\ft}  \nonumber\\
& = \frac{\ft\exp(-\ft^2/4)+(1+\ft^2/2)\sqrt{\pi}\left(1+\erf(\ft/2)\right)}
{2\exp(-\ft^2/4)+\ft\sqrt{\pi}\left(1+\erf(\ft/2)\right)}\;.
\end{align}

The large-size limit, $L\rightarrow \infty$ can be obtained in two ways, depending on whether the force is tensile (positive) or compressive (negative). Whether this is a thermodynamic limit or not depends on whether it results in uniform response.
From the definition of $\ft$, it is clear that the large tensile force ($\ft \gg 1$) limit coincides with the large chain limit ($L\gg l_p$) and yields a linear force-extension relation,
\begin{align}
\langle \tilde{z} \rangle =\frac{\ft }{2}\;,
\end{align}
which can be rewritten  as 
\begin{align}
\big\langle \frac{z}{L} \big\rangle =\frac{2 f l_p}{3 k_B T}\;.
\end{align}
The large-size limit in the  compressive case is obtained when $\ft \ll -1$ which coincides with the  limit $L\gg l_p$.  From the asymptotic series expansion of the error function for large negative $\ft $, 
\begin{align}
\label{asymptotic}
\erf\big(\frac{\ft }{2}\big)\approx -1+
\frac{1}{\sqrt{\pi}}\exp\big(-\frac{{\ft }^2}{4}\big)\big(-\frac{2}{\ft}+\frac{4}{{\ft }^3}-\frac{24}{{\ft }^5}+...\big)\;,
\end{align}
we obtain
\begin{align}
 \langle \tilde{z} \rangle \approx \frac{-8/{\ft }^3}{4/{\ft }^2}=-\frac{2}{\ft }\,
\end{align}
which is rewritten as
\begin{align}
\label{force-extension-Gibbs-compress}
\langle z \rangle =-2 \frac{k_B T}{f}\;.
\end{align}
We notice that Eq. (\ref{force-extension-Gibbs-compress}) is independent of the polymer size and therefore appears to pertain to the thermodynamic limit. Comparison, however, with Eq. (\ref{force-extension-Helmholtz-compress}) reveals a discrepancy by a factor of $2$ in the force-extension relations of the two ensembles in the compressional strong-force or long-chain limit. As with Eq. (\ref{force-extension-Helmholtz-compress}), we can recast  Eq. (\ref{force-extension-Gibbs-compress}) as a force-relative extension relation, involving only intensive (tension, relative extension) quantities,
\begin{align}
\label{force-strain-Gibbs-compress}
\big \langle \frac{z}{L}\big  \rangle =-2 \frac{k_B T}{Lf}\;.
\end{align}
In this form, the size ($L$) dependence persists,  and we conclude, as in the isometric ensemble, that the only bona fide thermodynamic limit exists for a tensile ($f>0$) force. 
%We also point out that, in both ensembles, in the compressive regime, the limits $f\rightarrow 0$ (or $\langle f \rangle \rightarrow 0$) and $L\rightarrow \infty$ do not commute.

As the discrepancy by a factor of $2$ between Eq. (\ref{force-strain-Gibbs-compress}) and Eq. (\ref{force-strain-Helmholtz-compress}) is a central result of this paper, it warrants further discussion. We can understand it by pointing out that in both ensembles, in the case of a Gaussian chain, the force (fixed or average) is purely entropic. As such, it is given by the derivative of the corresponding entropy with respect to the extension (fixed or average). As we show below, the discrepancy can be traced to the difference in the corresponding entropy. Let us first consider the Helmholtz ensemble. For strong compression, the relevant part of the Rayleigh distribution is the linear one (close to the origin). So, the entropy is $S_H\approx k_B \ln( z)$ (up to an irrelevant constant) and the (average) entropic force is\\
\begin{align}
\label{entropic_Helmholtz}
\langle f \rangle =-T\frac{\partial S_H}{\partial z}= -\frac{k_B T}{z}\;.
\end{align}
In the Gibbs ensemble, the corresponding entropy can be extracted from the partition function of Eq. (\ref{isotensional_Z_1}) which acts as a statistical weight.  Using the asymptotic series expansion of the error function, Eq. (\ref{asymptotic}), we obtain (in the large compressional force limit), $S_G\approx2k_B\ln(|f|)$ up to an irrelevant constant. Inserting the force from Eq. (\ref{force-extension-Gibbs-compress}) (notice that the prefactor of 2 in that equation is irrelevant for this particular step, so the argument is not circular!), we obtain $S_G\approx2k_B\ln(\langle z \rangle)$ up to an irrelevant constant, and
\begin{align}
\label{entropic_Gibbs}
 f  =-T\frac{\partial S_G}{\partial \langle z \rangle}=-2\frac{k_BT}{\langle z \rangle}\;.
\end{align}

\begin{figure}
 \centering
 \includegraphics[width=0.5\textwidth]{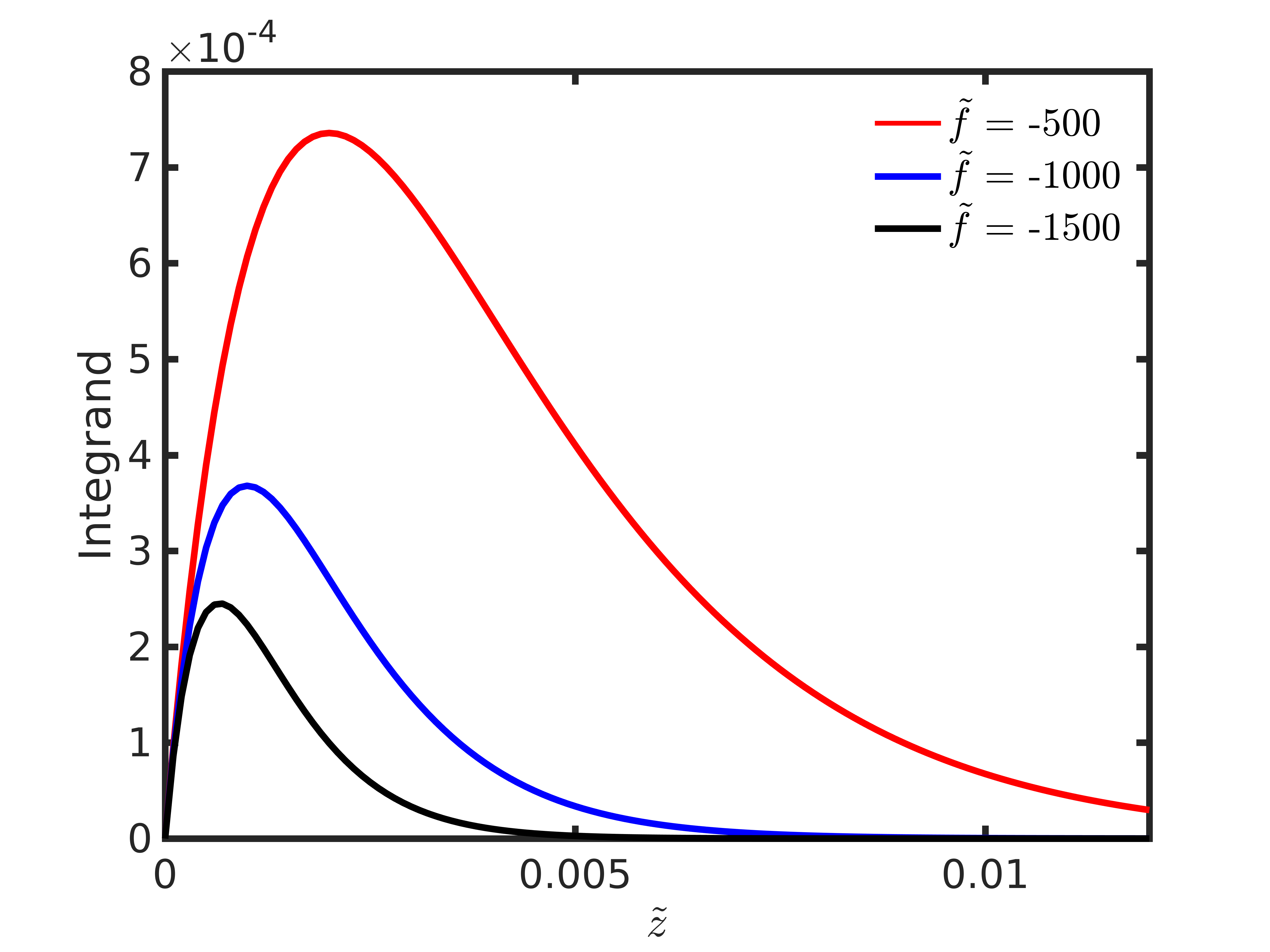}
 \caption{We plot the integrand of the partition function of Eq. (\ref{isotensional_Z_1}) which, up to a normalization prefactor, is the probability distribution of the free end in the isotensional ensemble, for three values of the compressive force. The skewness of the curve makes the average extension to always lie to the right of the extension which corresponds to the peak. The latter is the extension of the isometric ensemble.}
 \label{fig:integrand}
\end{figure}

We gain more insight into the origin of this discrepancy, if we plot the probability distribution of the free end for large compressional forces (Fig \ref{fig:integrand}).  This probability distribution is given by the integrand of the second line of Eq. (\ref{isotensional_Z_1}) (up to a normalization prefactor). The maximum of the probability corresponds to an extension $z_{\rm max}$ which is exactly what we get from the Helmholtz ensemble, Eq. (\ref{force-extension-Helmholtz}). However, because of the skewness of the distribution, the average extension is always to the right of $z_{\rm max}$: $\langle z \rangle > z_{\rm max}$. This is an effect of the confinement due to the substrate. There is no way to approximate the distribution by a Gaussian with non-zero average, because the standard deviation is significant down to the limit of infinite compressional force. The variance of the distribution is \\
\begin{align}
\label{variance}
\langle z^2 \rangle-\langle z\rangle^2=k_B T\frac{\partial\langle z\rangle}{\partial f}\;.
\end{align}
The standard deviation,  $\sigma_z=\sqrt{\langle z^2 \rangle-\langle z\rangle^2}$, for strong negative forces becomes $\sigma_z=\langle z\rangle/\sqrt{2}$ (cf. Eq. (\ref{alphaG})).  We see that it remains of the order of the average displacement which implies that fluctuations below the average (compressive) are always affected by the substrate, whereas fluctuations above the average (stretching) are not. That is why the saddle-point approximation of the second line of Eq. (\ref{isotensional_Z_1}) breaks down and fails to yield the average $z$. Notice that the ratio $\sigma_z/\langle z \rangle =1/\sqrt{2}$ is finite and independent of the size of the system. This is a clear marker of the non-existence of the thermodynamic limit which is always behind the ensemble inequivalence. We should contrast this to the Gaussian chain in free space which has $\sigma_z/\langle z \rangle=k_BT\sqrt{3}/\sqrt{2Ll_p}$ and vanishes as the size $L$ increases. (In the Helmholtz ensemble, for $z$ in the strongly compressive regime, we get from Eq. (\ref{force-extension-Helmholtz-compress})  $\sigma_f/\langle f \rangle=1$ which again shows significant fluctuations relative to the average, independent of the size of the system.)

It is interesting to point out that one way to reach the thermodynamic limit where the two curves coincide is to decrease the temperature of the tensile (positive force) system, keeping everything else fixed. In both ensembles, the temperature appears only in the denominator of our dimensionless force, $\ft$. We can understand this behavior as resulting from a decrease in the frequency of collisions with the substrate due to a decrease in thermal fluctuations. As the role of the substrate becomes less significant, we approach the thermodynamic limit of the Gaussian chain in free space.

\begin{figure}
 \centering
 \includegraphics[width=0.5\textwidth]{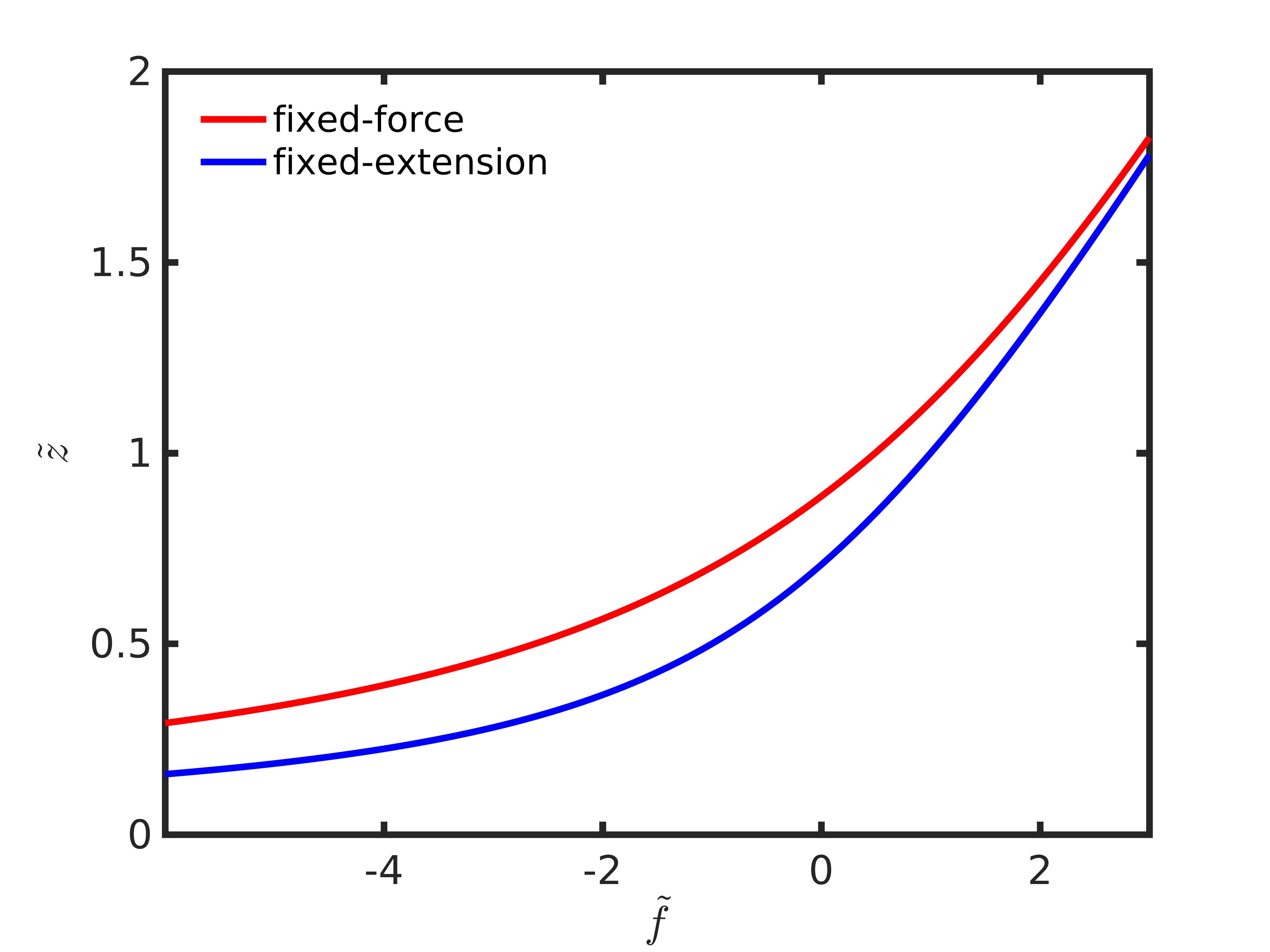}
 \caption{Force-extension relation in fixed-force and fixed-extension ensemble.  The dimensionless force, $\tilde{f}$, and the dimensionless extension, $\tilde{z}$, are defined in Eq. (\ref{force-extension-Helmholtz}). We display the two curves over  the range of parameters where the ensemble inequivalence is clearly pronounced. }
 \label{fig:gaussian}
\end{figure}

\begin{figure}
 \centering
 \includegraphics[width=0.5\textwidth]{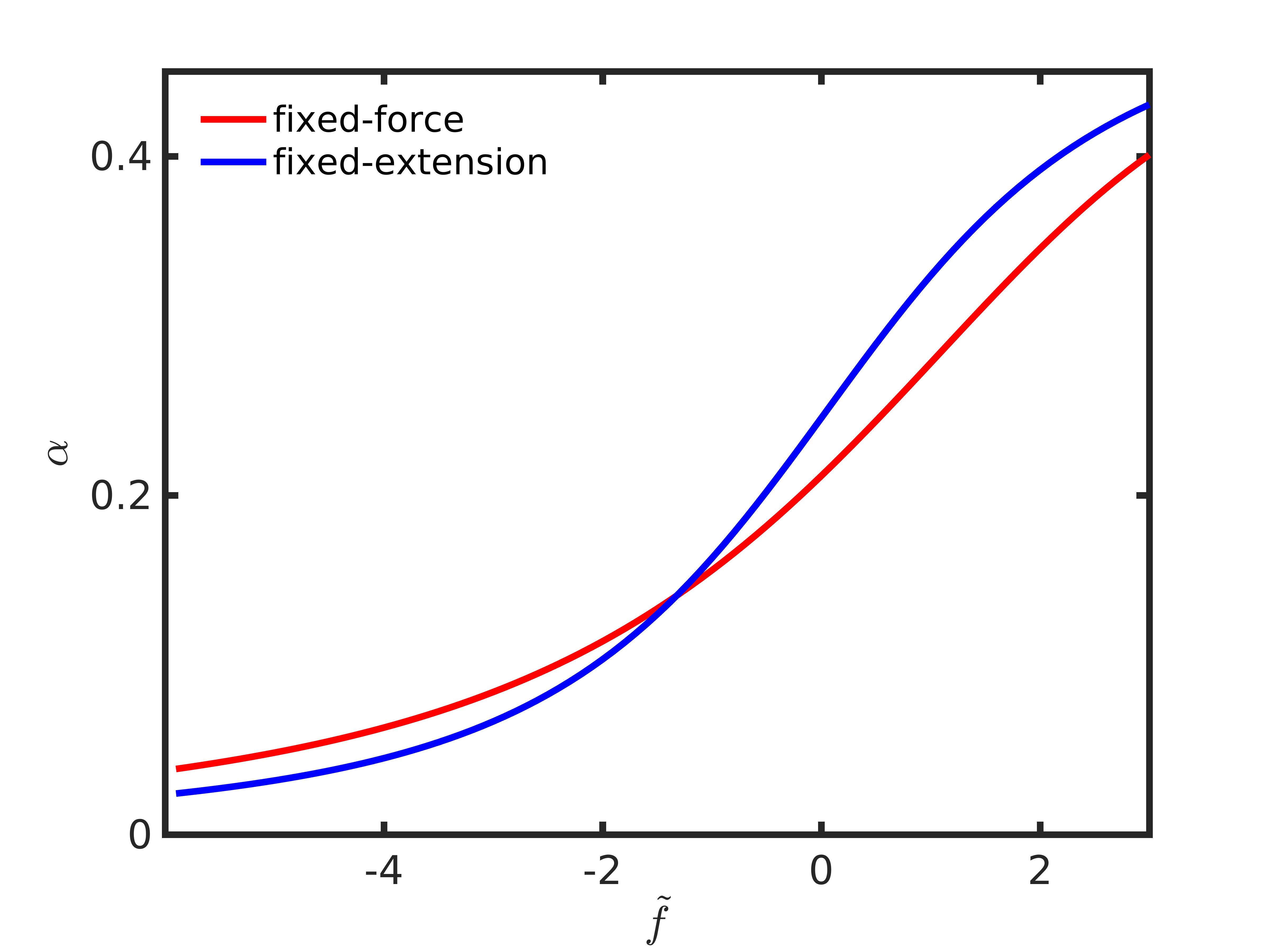}
 \caption{A measure of  the elasticity of a prestressed polymer is the differential  compliance as defined in Eqs. (\ref{complianceG}) and (\ref{complianceH}) for the two ensembles. The plot shows that even  though the force-extension curves look similar for both  ensembles at large positive forces, their compliance is very different. $\alpha$ is measured in units of $4l_p/(3K_BT)$}.
 \label{fig:gaussian_compliance}
\end{figure}

% Expressions for $\delta z$
% from Eq 9 seems to reproduce Eq 8 quite well as shown in Fig \ref{fig:comparison}
% below
% \begin{figure}
%  \centering
%  \includegraphics[width=0.6\linewidth]{codes/force_ensemble/dz.jpg}
%  \label{fig:comparison}
% \end{figure}

\section{Differential compliance (susceptibility) of the tethered chain}

The ideal Gaussian chain exhibits linear elasticity (entropic spring with a well-defined spring constant) as shown in Sec. 2. More realistic polymer models (freely-jointed chain, wormlike chain, etc.) have non-linear force-extension relations. As we saw in the previous sections, even an ideal Gaussian chain exhibits nonlinear tensile and compressive elasticity when it is tethered to a planar rigid substrate. A very useful measure of the elastic response of a non-linear system which can be extracted from the force-extension relation is the differential compliance. It is the change in the polymer extension that corresponds to a small incremental change in its tensile (positive) or compressive (negative) force. In the Gibbs ensemble, it corresponds to the susceptibility in statistical physics (cf. Eq. (\ref{variance})). In the Helmholtz ensemble, the susceptibility corresponds to the differential stiffness which is the inverse of the differential compliance. More precisely, the differential compliance in the Gibbs ensemble is defined as\\
\begin{align}
{\alpha}_G=\frac{1}{L}\frac{\partial \langle z \rangle}{\partial f}\;,
\label{complianceG}
\end{align}
whereas in the Helmholtz ensemble it is
\begin{align}
\alpha_H=\frac{1}{L}\frac{\partial z }{\partial \langle f \rangle}\;.
\label{complianceH}
\end{align}
In case of a Gaussian chain tethered to a wall, the differential compliance can be calculated exactly in each ensemble using Eqs. (\ref{force-extension-Gibbs}) and (\ref{force-extension-Helmholtz}),
\begin{widetext}
\begin{align}
\tilde{\alpha}_G & = \frac{4e^{-\frac{\ft^2}{2}}+3\sqrt{\pi}\ft e^{-\frac{\ft^2}{4}}\left(1+\erf(\frac{\ft}{2})\right) + \pi\left(-1+\frac{\ft^2}{2}\right)\left(1+\erf(\frac{\ft}{2})\right)^2}{\left(2e^{-\frac{\ft^2}{4}}+\sqrt{\pi}\ft\left(1+\erf(\frac{\ft}{2})\right)\right)^2}\label{alphaG} \\
\tilde{\alpha}_H & = \frac{1}{4}\left(1 + \frac{\ft}{\sqrt{{\ft}^2+8}}\right), \label{alphaH}
\end{align}
\end{widetext}
where $\tilde{\alpha}=\alpha 3 k_BT/(4l_p)$.

In Fig. \ref{fig:gaussian_compliance}, we plot the differential compliance in the two ensembles, where we use the rescaled dimensionless force $\tilde{f}$. We notice that the two ensembles give different values for the compliance except for a point where the two force-extension curves appear to be close to their maximal difference. In the tensile (positive force) regime, the chain is softer in the Helmholtz ensemble, while in the compressive (negative force) regime, it starts softer in the Helmholtz ensemble and, as the negative force increases, it becomes softer in the Gibbs ensemble. It is remarkable, that for relatively large extension (fixed or average), close to $0.5 L$, where the ensemble difference of the two force-extension curves seems negligible, the corresponding relative difference in the differential compliance is significant. In the strong tensile limit $\ft>>1$, in both  ensembles, $\tilde{\alpha}_G$ and $\tilde{\alpha}_H$ tend to the constant compliance of a free Gaussian chain $ \rightarrow\frac{1}{2}-\frac{1}{{\ft}^2}$.  However, in the strong compression limit, $\ft << -1$, their inequivalence shows up in the rates at which they vanish: $\tilde{\alpha}_G\rightarrow\frac{4}{\ft^2}$ and $\tilde{\alpha}_H\rightarrow\frac{1}{\ft^2}$, as expected from the force-extension relations.

%Of course, in the thermodynamic limit of very large positive force (fixed or average), we expect the corresponding values of the differential compliance in the two ensembles to converge to the universal value of $2 l_p/(3 k_B T)$. In the \textcolor{blue}{opposite limit of infinite compressional force}, the differential compliance vanishes.

\section{Tethered wormlike chain}

\begin{figure}
\centering
\includegraphics[width=0.5\textwidth]{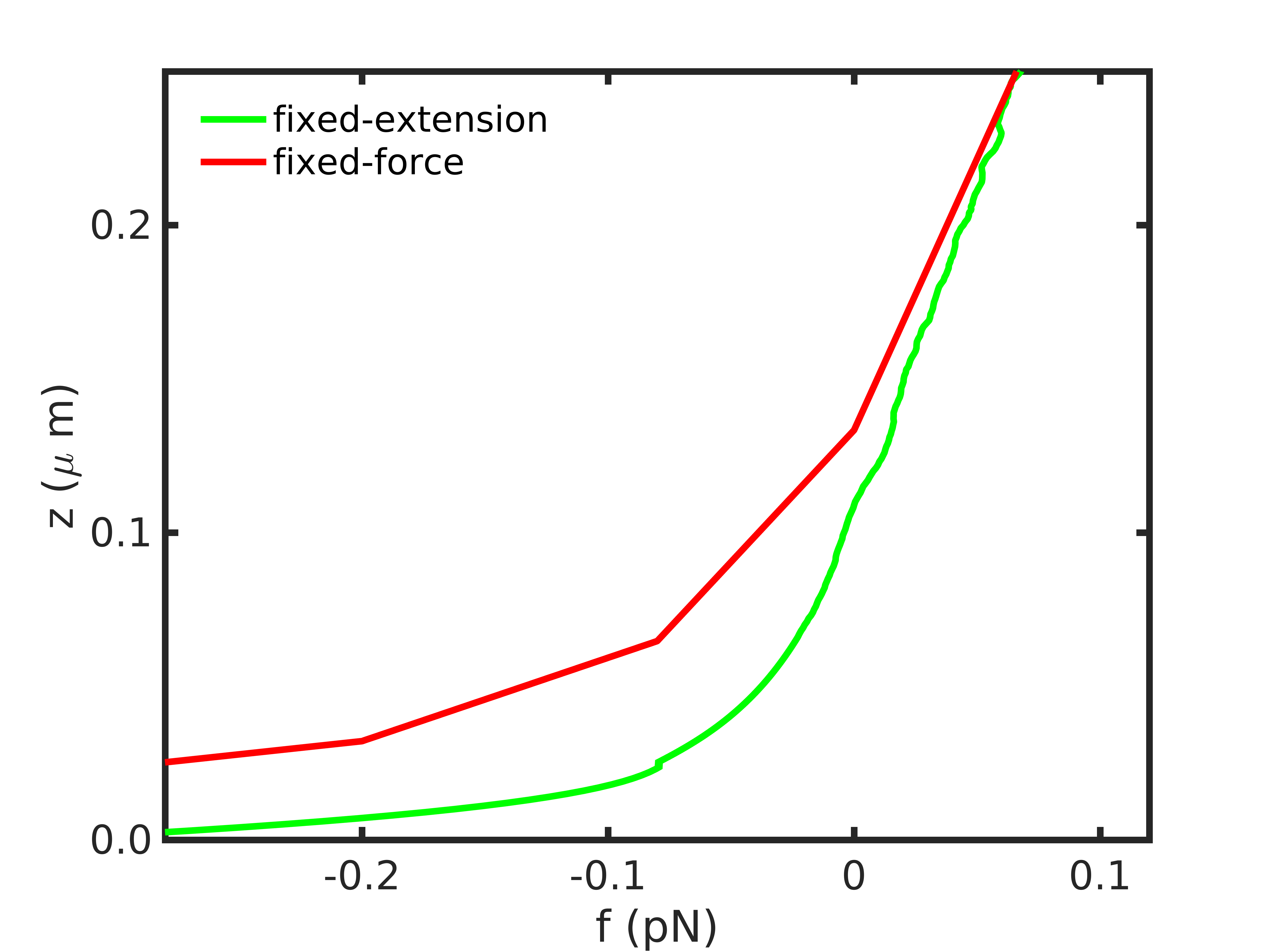}
\caption{Force-extension relations in the two ensembles obtained by Monte Carlo simulations for a tethered wormlike chain with parameters corresponding to a short double-stranded DNA ($L=1 {\rm \mu m}$ and $l_p= 50 {\rm nm}$) }
\label{fig:wlc}
\end{figure}

As opposed to synthetic polymers (e.g., polyethylene), biopolymers (including DNA) exhibit semiflexibility, that is, their conformations are dominated by their bending stiffness and their (approximate) local inextensibility. A widely used minimal theoretical model which captures those features is the wormlike chain (WLC) \cite{STY}. It is a one-dimensional fluctuating object with local inextensibility and a finite bending stiffness. Motivated by the experimental study of force-free conformations of a tethered short DNA chain by Lindner et al. \cite{Lindner}, we model the DNA chain as a WLC and study the two ensembles numerically using Monte-Carlo simulations. First, we look at the fixed-force ensemble for a chain tethered at one end to an impenetrable wall (Fig. \ref{fig:wall}-(b)), and pulled at the other end by a force $\mathbf{f}$. The  chain is of total contour length $L$ and bending stiffness $\kappa$. Its elastic Hamiltonian is 
\begin{equation}
\mathcal{H} = \frac{\kappa}{2}\int_0^L ds\left(\frac{\partial\mathbf{t}(s)}{\partial s}\right)^2 - \mathbf{f}\cdot\int_0^Lds\mathbf{t}(s)\;,
\label{wlccontinuous}
\end{equation}
where $\mathbf{t}(s) = d\mathbf{r}(s)/d s$ is the tangent vector of the space curve $\mathbf{r}(s)$ parameterized by the arc-length position $s$. In addition, the polymer is locally inextensible which is mathematically imposed by $\vert\mathbf{t}(s)\vert = 1$. For the simulation, we use so called Kratky-Porod model \cite{KratkyPorod} which is the   discretized version of the WLC. When discretized into $N = L/\Delta s$ segments of size $\Delta s$, the Hamiltonian becomes the same as that of \citet{lattanzi2004transverse,gholami2006entropic},
\begin{equation}
\mathcal{H} = -\frac{\kappa}{\Delta s}\sum_{i=1}^{N-1}\mathbf{t}_i\cdot \mathbf{t}_{i+1} - \Delta s\sum_{i=1}^N\mathbf{f}\cdot \mathbf{t}_i\;.
\label{wlcdiscrete1}
\end{equation}

Since the translational invariance is broken due to the wall,  it becomes prohibitively difficult to show the inequivalence of the two ensembles  theoretically. We use Monte Carlo simulations with the Kratky-Porod Hamiltonian  in Eq. (\ref{wlcdiscrete1}) to obtain the force extension relation. The effect of a tethering wall on the force-extension relation (in the fixed-force ensemble) of a helical semiflexibe filament \cite{PanyukovRabinPRE2000} has previously been investigated with Monte Carlo simulations by Kessler and Rabin \cite{KesslerRabinJCP2004}.  A similar Monte Carlo scheme has been used by Blundell and Terentjev \cite{BlundellEMJSoftMatter2011} to compute the force-extension relation of such a filament in the presence of steric constraints, albeit focusing on confinement by cylindrical walls.

We use the parameters of the experiment in Ref. \cite{Lindner}: total contour length $L=1{\rm \mu m}$ and persistence length $l_p=50 {\rm nm}$. Details of the simulations  are given in the Appendix \ref{appendix}.  For the constant extension ensemble, we follow the standard procedure for the calculation of entropic forces, as we did for the Gaussian chain in Eq. (\ref{entropicforceGaussian}) : first we obtain the conformational distribution of the free end with no external force, and then we take its derivative with respect to the axial ($z$) displacement to obtain the corresponding component of the entropic force.  In the constant force ensemble, we simulate the average extension with a constant applied force. The force-extension results are different for the two ensembles as shown in Fig. \ref{fig:wlc}.  We see that the ensemble inequivalence starts to become significant for rather low tensile forces, less than $\sim 0.08 {\rm pN}$. Seol et al. \cite{SeolPNelsonBPJ2007} measured the force-extension relation of short ($0.6-7 {\rm \mu m}$) tethered DNA molecules using an optical trap, and found that below $0.08 {\rm pN}$ the presence of the wall becomes significant, primarily because of bead-substrate excluded-volume interactions. This is confirmed in subsequent experiments by te Velthuis et al. \cite{teVelthuisBPJ2010} and analysis by Mehraeen and Spakowitz \cite{tether_Spakowitz}. $0.08 {\rm pN}$ is a characteristic force scale for double-stranded DNA, because it is equal to the ratio $f_T:=k_BT/l_p$ of the typical thermal energy over the persistence length. It is known that such small forces are crucial in protein-mediated loop formation \cite{femtonewtonMeinersPRL2010} and can be probed experimentally with axial optical tweezers \cite{MeinersJoVE2011}.

The strong stretching response of a free (that is, in the absence of any confinement) WLC is given by the Marko-Siggia force-extension formula \cite{marko1995stretching},
\begin{align}
\label{Marko-Siggia}
\Big\langle \frac{z}{L} \Big \rangle = 1-\frac{k_B T}{\sqrt{4f\kappa}}\;,
\end{align}
which holds for $f\gg \kappa/(l_p^2)=f_T$ (so that $\langle z/L\rangle\approx 1$). This condition is equivalent to $l_p \gg \sqrt{\kappa/f}$. It is known that $\sqrt{\kappa/f}$ is the length of propagation of boundary effects along the polymer contour \cite{PGDG_Ciferri,PBEMJ}. If it is much less than the total contour length $L$, the stretched WLC behaves independently of the boundary conditions \cite{PBHinge,PBMotor} and therefore it is in the thermodynamic limit. This can be seen in Eq. (\ref{Marko-Siggia}), because it is invariant to a rescaling of the extensive variables $L$ and $z$. 

The axial probability distribution of the free end which is needed in the isometric ensemble is shown in Fig. \ref{fig:wlc_free_end} in the Appendix. It agrees with a similar simulation in Ref. \cite{Lindner} which showed a shift towards the substrate of the WLC compared to a Gaussian chain with the same parameters. That discrepancy is due to the finite bending stiffness of the former, which biases conformations with small slope from the substrate after a collision with it.

\section{\label{conclusion}Summary-Conclusions} 

In this paper, we analyzed the force-extension relation in the fixed-extension and the fixed-force ensemble for an ideal (Gaussian) chain with one end tethered to a rigid planar wall. We obtained closed analytic expressions for the two relations. Their difference establishes the inequivalence of the two ensembles which is caused by the tethering substrate. We emphasize that this is purely a confinement effect and does not involve any potential interaction. Whereas for a free Gaussian chain we have a linear force-extension relation in both ensembles, tethering to a substrate yields nonlinear relations. It turns out that, for a given force (fixed or average), the corresponding extension (average or fixed, respectively) is larger in the isotensional (Gibbs) ensemble. This difference is understood from the skewness of the probability distribution of the extension in the isotensional ensemble (Fig. \ref{fig:integrand}). We have shown that the value of the extension that corresponds to the peak, $z_{\rm max}$, is precisely equal to the extension in the isometric ensemble. Because of the skewness, we get $z_{\rm max}<\langle z \rangle$. For large tensile forces, the two curves asymptotically converge to a single linear relation. This is to be expected, because in that limit, the role of the substrate gets diminished. In the opposite limit of large compressive forces, the corresponding extension tends to zero hyperbolically in both ensembles. The respective relations, however, differ by a factor of $2$.  This significant discrepancy is a major result of our article. It can be understood by the persistence, in the strongly compressional limit, of significant fluctuations relative to the average which do not depend on the size of the polymer. This implies the non-existence of a uniform thermodynamic limit which, in turn, becomes manifest as inequivalence of ensembles. The sensitivity of the force-extension relations  to the size of the polymer is clear. If we express the response in terms of force and relative extension (defined as the ratio of the extension to the size of the polymer) and take the thermodynamic limit of large polymer size, we obtain the  linear relations of strong tensile forces. For compressive forces, the relative extension in the strong-force or, equivalently, long-chain limit reduces to zero. In the strongly compressive regime, the size dependence of the force-extension relations drops out, but the ensemble inequivalence persists due to the factor of $2$. Apart from the force-extension relations, we calculated the differential compliance of the tethered chain in the two ensembles. In the tensile regime, the differential compliance is larger for the isometric ensemble. Remarkably, the ensemble difference of the differential compliance does not follow quantitatively the ensemble difference of the force-extension relation. The former can be significant where the latter is negligible and vice versa. 

In order to test the relevance of our predictions to experiments involving tethered DNA, we also did Monte Carlo simulations of the wormlike chain (WLC) model with contour length $L=1\mu m$ and persistence length $l_p=50 nm$, parameters used in the experiment of Ref. \cite{Lindner}. The ratio $L/l_p=20$ implies a fairly flexible chain. As such, the force-extension curves qualitatively agree with those of the Gaussian chain, exhibiting ensemble inequivalence with larger extension in the fixed-force ensemble. However, we also confirm the finding of Ref. \cite{Lindner} about the average axial extension of the WLC being smaller for the WLC compared with that of the Gaussian chain. That is understood because of the finite bending stiffness of the latter which biases the direction of the tangent vector after a collision with the substrate not to deviate much from the plane. We point out that the ensemble inequivalence shows up for forces less than  $0.08{\rm pN}$  which is a characteristic value for the strong (global) bending of DNA. It is therefore caused by the confinement due to the tethering substrate. We have shown that the effect of the confinement is qualitatively the same in the Gaussian chain model, in the WLC model, and in the bond fluctuation (BF) model of Ref. \cite{JuttaBinderJCP2014}.

In our study, a force is applied or measured at the free end (a single point)  of a tethered polymer. In many experimentally relevant situations, however, the polymer interacts with an extended surface which acts like a piston. That surface may represent, for example, the finite extent of an AFM probe.  Kantor and Kardar \cite{KantorKardarPRE2017} have calculated the entropic force due to the confinement of a polymer between scale-free surfaces (e.g., wedges, cones, flat plates). Edwards and Freed \cite{EdwardsFreed_eqofstate} have calculated the equation of state (pressure-volume relation) of an ideal (Gaussian) chain confined in a rectangular box. Those studies are done in the isometric ensemble. The corresponding calculations in the isotensional ensemble remain an open problem which would be an interesting extension of our present work.

\appendix
\section{\label{appendix} Simulations}
\begin{figure}
\centering
\includegraphics[width=0.5\textwidth]{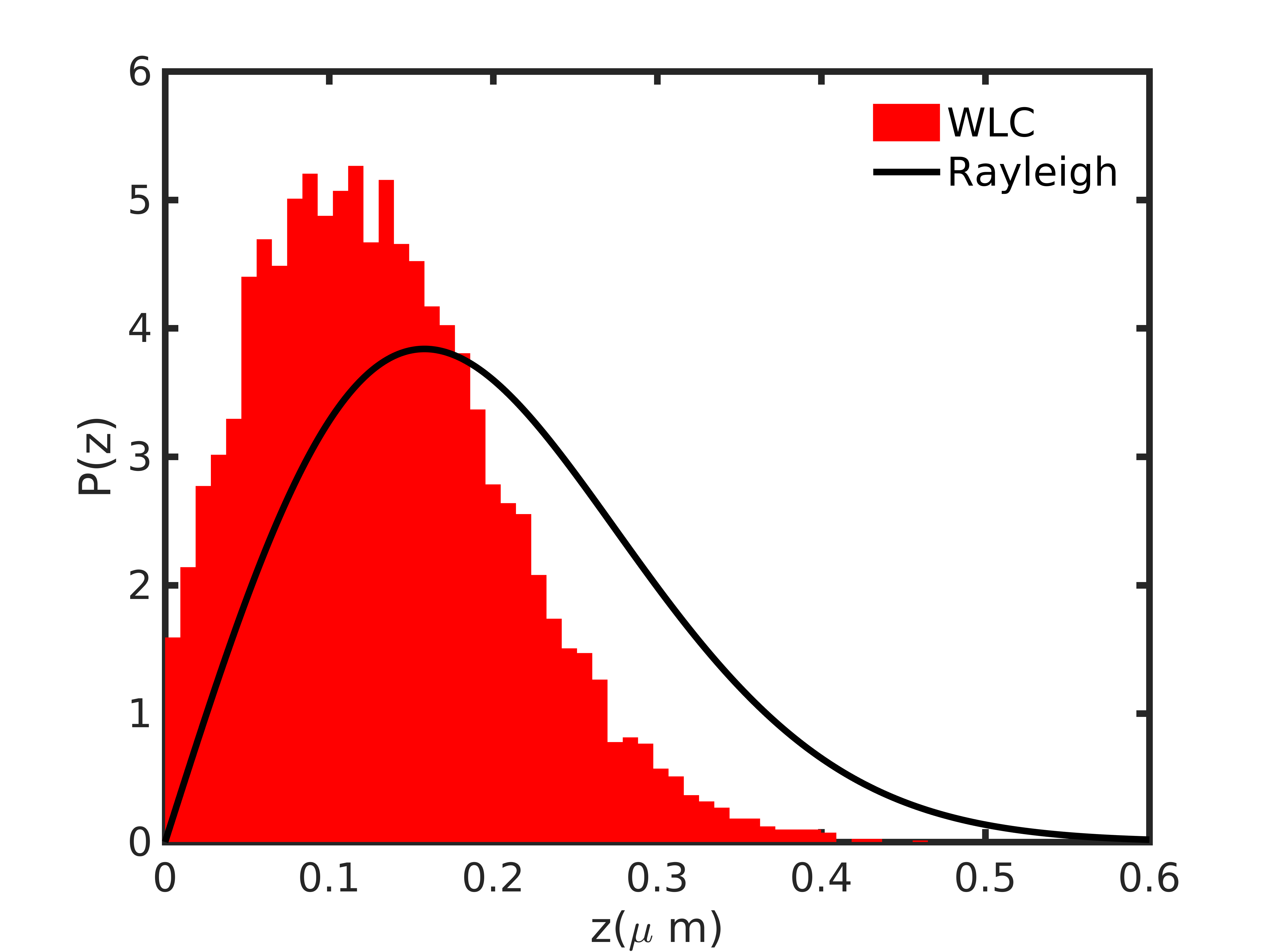}
\caption{The distribution of the free end probability of a wormlike chain tethered to a wall at $z = 0$.  The deviation from the corresponding Rayleigh distribution of a Gaussian chain is a result of the bending rigidity of the wormlike chain. }
%The derivative of the distribution as in Eq (\ref{entropicforceGaussian}) gives the force extension in fixed-extension ensemble in Fig \ref{fig:wlc}.
\label{fig:wlc_free_end}
\end{figure}

We consider a short DNA chain of length $L = 1{\rm \mu}$m and persistence length $l_p = 50$ nm in Eq.(\ref{wlcdiscrete1}) ($k_B T=4.1$ pN nm at $24^{\circ}$C). The chain is discretized into $1000$ segments of length $\Delta s = 1$nm. For the fixed-force ensemble, the Hamiltonian in Eq. (\ref{wlcdiscrete1}) is simulated with the Monte Carlo algorithm of Ref. \cite{gholami2006entropic}. The system is averaged over $\sim 10^6$ time steps to equilibration. For the fixed-extension ensemble, we first calculate the probability distribution of the free end of the chain as shown in Fig. \ref{fig:wlc_free_end}. The entropic force in Fig. \ref{fig:wlc} is obtained by taking a numerical derivative as prescribed in Eq. (\ref{entropicforceGaussian}).  

%If notes are included in your references you can change the title from 'References' to 'Notes and references' using the following command:
%\renewcommand\refname{Notes and references}

%%%REFERENCES%%%
\bibliography{bibliography} %You need to replace "rsc" on this line with the name of your .bib file

\end{document}